\documentstyle{article}
\def\dbox#1#2{\mbox{\fbox{$\begin{array}{c}#1\\ #2\end{array}$}}}
\def\darrow{\Big\downarrow}
\def\rarrow{\!\rightarrow\!}
\def\ti{\tilde}
\def\ve{\varepsilon}

\def\bb{\begin{equation}}
\def\ee{\end{equation}}
\def\th#1{\vspace{1mm}\noindent{\bf #1}\hspace{2mm}}
\def\proof{\noindent {\bf Proof.}\hspace{2mm}}
\def\qed{\vrule height5pt width4pt depth0pt \vspace{1mm}}
\def\bitem#1#2#3#4#5#6{\bibitem{#1}{#2 #3 {\it #4} {\bf #5} #6}}

\begin{document}
\begin{titlepage}
\title{Integrable boundary conditions for the Toda lattice}
\author{V.E. Adler and I.T. Habibullin                            \\
        Ufa Institute of Mathematics, Russian Academy of Sciences,\\
        Chernyshevsky str. 112, 450000 Ufa, Russia                \\
        e-mail: adler@nkc.bashkiria.su, ihabib@nkc.bashkiria.su}
\date{\today}
\maketitle
\thispagestyle{empty}
\vspace{5mm}
\begin{abstract}
The problem of construction of the boundary conditions for the Toda lattice
compatible with its higher symmetries is considered.  It is demonstrated that
this problem is reduced to finding of the differential constraints consistent
with the ZS-AKNS hierarchy.  A method of their construction is offered based
on the B\"acklund transformations.  It is shown that the generalized Toda
lattices corresponding to the non-exceptional Lie algebras of finite growth
can be obtained by imposing one of the four simplest integrable boundary
conditions on the both ends of the lattice.  This fact allows, in particular,
to solve the problem of reduction of the series $A$ Toda lattices into the
series $D$ ones.  Deformations of the found boundary conditions are presented
which leads to the Painlev\'e type equations.

\vspace{5mm}
Key words: Toda lattice, boundary conditions, integrability, B\"acklund
transformation, Lie algebras, Painlev\'e equations
\end{abstract}
\end{titlepage}

\section{Introduction}

In this paper we consider the boundary conditions for the classical
completely integrable differential-difference model --- the well known Toda
lattice
\bb q_{j,xx}= \exp(q_{j+1}-q_j)-\exp(q_j-q_{j-1}),\;\; -\infty<j<+\infty
\label{toda}\ee
which are compatible with the integrability property.  Imposing of the
boundary conditions of the form
\bb q_M=F(q_{M+1},\dots, q_{M+m}), \;
    q_N=G(q_{N-1},\dots, q_{N-n}), \; M<N \label{bound}\ee
reduces the infinite-dimensional system (\ref{toda}) to a finite-dimensional
one.  We require that the boundary conditions (\ref{bound}) are consistent
with integrability at any choice of parameters $M<N$ (for fixed $m,n$).
Thus, the boundary conditions on the left and right ends are assumed
independent from each other, so actually we work with a semi-infinite
lattice.

The compatibility of a boundary condition with the integrability property is
understood, in spirit of the symmetry approach, as consistence with the flows
determined by higher symmetries of the Toda lattice.  (The exact definition
is given below, in Section 2.) Such interpretation was suggested in the
papers \cite{hab1},\cite{gur1} and has quite justified itself in the problem
of description of the integrable boundary conditions for such models, as
Harry Dym and Korteweg-de Vries equations, multifield systems of the NLS and
Burgers type, Volterra lattice and others (see \cite{hab2}, \cite{gur2},
\cite {hsvi}).

In some cases (for example, for the Burgers equation see \cite{gur2}) one can
prove that a boundary condition compatible at least with one higher symmetry
of the equation is compatible with an infinite series of its symmetries as
well, including a certain one, which can easily be specified a priori and
consequently may be regarded as a test symmetry.  Compatibility with the test
symmetry is taken as a basis of the preliminary classification of the
integrable boundary conditions (see Section 3).

The finite-dimensional versions of the lattice (\ref{toda}) called
generalized Toda lattices are well known in literature.  Namely, to each
simple Lie algebra of the finite growth the generalized Toda lattice
corresponds which is integrable in some sense (see e.g \cite{lez},
\cite{bog}).  In this paper we undertake an attempt to revise the theory of
the finite-dimensional integrable systems of exponential type from the
integrable boundary conditions point of view.  For example, we demonstrate
that the generalized Toda lattices corresponding to the classical series of
finite-dimensional simple Lie algebras, as well as Kac-Moody algebras, can be
represented as reductions of the infinite lattice (\ref{toda}) with the
boundary conditions (\ref{bound}) compatible with its higher symmetries.  It
should be emphasized that the periodic lattices corresponding to the series
$\tilde A,$ as well as the lattices related to the exceptional Lie algebras,
remain outside of our consideration by virtue of the assumption about
independence of the boundary conditions on the left and right ends.  On the
other hand, the approach developed in the work makes possible to study also
boundary conditions leading beyond the limits of exponential type systems
(see, for example, boundary condition (\ref{b2}) with a complete set of
parameters).  The revealing of the algebraic structures appropriate such
boundary conditions is an open problem.

The article is organized as follows.  In Section 2 the compatibility
criterion of a boundary condition with a higher symmetry of a lattice is
formulated.  It involves the dual language of evolutionary partial
differential equations associated to the lattice.  In this Section we discuss
also a phenomenon of a degenerate boundary condition, which is characterized
by increased integrability.

In Section 3 we present a method to construct the integrable boundary
conditions for the Toda lattice which is based on this criterion.  Symmetries
of the Toda lattice, rewritten in the evolutionary form, coincide with the
ZS-AKNS hierarchy and finding of integrable boundary conditions is equivalent
to finding of differential constraints compatible with the odd members of
this hierarchy, in particular, with the system (\ref{ckdv}) which plays the
role of the test symmetry.  This problem is solved with help of the
B\"acklund transformations.  It is noticed that by virtue of the found small
order differential constraints the system (\ref{ckdv}) is reduced into the
integrable scalar equations KdV, mKdV and Calogero-Degasperis equations.

In Section 4 we demonstrate that imposing of the every possible combinations
of the four simplest boundary conditions found above on the left and right
ends of the Toda lattice leads to the generalized Toda lattices,
corresponding to all infinite series of the Lie algebras of the finite growth
(except for the $\ti A$ series).  Moreover, it is proved that the series $D$
Toda lattices can be represented as reductions of the series $A$ lattices.
We shall note, that though for the lattices of the series $B$ and $C$ the
connection with the series $A$ is quite obvious and well known, for the
lattices of the series $D$ this problem remained open.\footnote{Authors thank
R.I.~Yamilov who drew their attention to this problem.}  The propositon 4.1
is proven:  let the solution of the Toda lattice corresponding to the Lie
algebra $A_{2n-1}$ satisfies an additional symmetry of the reflection type,
then it sets also solution of the Toda lattice corresponding to the Lie
algebra $D_n.$

In Section 5 the found boundary conditions are deformed by transform
$q_j\rightarrow q_j+\ve x.$  The new boundary conditions are also integrable,
but they are compatible with other symmetry set.  We demonstrate that closing
of the lattice (\ref{toda}) by the boundary conditions with different values
of parameter $\ve$ on the right and left ends results in equations of the
Painlev\'e type.

\section{Boundary conditions compatible with higher symmetries}

Let a lattice of the Toda type
\bb q_{j,xx}=f(q_{j-1},q_j,q_{j+1}),\; -\infty<j<+\infty  \label{chain}\ee
where ${\partial f\over\partial q_{j\pm 1}}\neq 0$
admits a higher symmetry of the form
\bb q_{j,t}=g(q_{j-k},q_{j-k,x},\dots,q_{j+k},q_{j+k,x}). \label{sym}\ee
Using the boundary condition
\bb q_0 = F(q_1,q_{1,x},q_2,q_{2,x}\dots,q_m,q_{m,x})     \label{bc}\ee
we shall reduce the lattice (\ref{chain}) to a semi-infinite lattice
determined on the right half-axis $j>0.$  Everywhere below we assume local
analyticity of the functions $f,g,F.$  Notice that, generally speaking, for
$k>1$ the boundary condition (\ref{bc}) is not sufficient to close the
lattice (\ref{sym}).  For this purpose in addition to (\ref{bc}) it is
necessary to express the variables $q_{-1},q_{-1,x},q_{-2},q_{-2,x},\dots,
q_{1-k},q_{1-k,x}$ through the dynamic variables $q_1,q_{1,x},q_2,q_{2,x}
\dots.$  Assuming that variables $q_0,q_{-1},\dots,q_{1-k}$ also satisfy
equation (\ref{chain}), one can easily obtain an algorithm of the boundary
condition (\ref{bc}) continuation.  First of all, differentiating (\ref{bc}),
one finds expression for $q_{0,x}.$  Then, solving the equation
$f(q_{-1},q_0,q_1)=D^2_x(F)$ with respect to $q_{-1},$ one obtains
$$q_{-1}=F^{(-1)}(q_1,q_{1,x}\dots,q_{m+1},q_{m+1,x}).$$
Further on each following step one finds $q_{-j}=F^{(-j)}(q_1, q_{1,x}\dots)$
from equation
$$f(q_{-j},F^{(1-j)},F^{(2-j)})=D^2_x(F^{(1-j)}).$$

\th{Definition.} The boundary condition (\ref{bc}) is called degenerate if
the following identity holds
$$
{\partial f(q_{-1},q_0,q_1)\over\partial q_{-1}}\Bigr\vert_{q_0=F}\equiv 0.$$

Obviously, for the degenerate boundary condition the continuation algorithm
is not valid.

\th{Proposition 2.1 } Let the boundary condition (\ref{bc}) is degenerate,
then $F$ does not depend on variables
$q_{1,x},q_2,q_{2,x},q_3,q_{3,x},\dots$ (but can depend on $q_1$).

\proof Assume that variables $q_1$ and $F$ are independent, i.e. $F$ depends
on $q_m$ or $q_{m-1,x}\; m>1.$  Then identity ${\partial f\over\partial
q_{-1}}(q_{-1},F,q_1) \equiv 0$ yields that the function
$f(q_{j-1},q_j,q_{j+1})$ does not depend on variable $q_{j-1}$ which
contradicts the form of the lattice (\ref{chain}).  \qed

\th{Consequence.} Let the boundary condition (\ref{bc}) is not degenerate,
then the algorithm of continuation given above is correct.

\proof In identity
${\partial f\over\partial q_{-j}}(q_{-j},F^{(1-j)},F^{(2-j)})\equiv 0$
the functions $F^{(2-j)}(q_1,\dots,q_{m+j-2})$ and
$F^{(1-j)}(q_1,\dots,q_{m+j-1})$ depend on the different sets of dynamic
variables and therefore are independent.  Hence, from this identity the
contradiction follows that the function $f$ in (\ref{chain}) essentially
depends on the third argument. \qed

\th{Proposition 2.2} Let the boundary condition (\ref{bc}) is degenerate,
then it completely closes the lattice (\ref{sym}) on the half-axis $j>0.$

\proof From the degeneracy of the boundary condition (\ref{bc}) it follows
that variables $q_{-1},q_{-1,x},$ $q_{-2},q_{-2,x},\dots$ can not be
expressed through the dynamic variables $q_1,q_{1,x},q_2,q_{2,x}\dots$ of the
problem (\ref{chain}),(\ref{bc}).  Therefore they can be considered as
independent ones.  From the condition of commutation of the flows determined
by equations (\ref{chain}), (\ref{sym}) in the point $j=1$ one obtains
$D_t(f_1)=D^2_x(g_1)$ where the subscript denotes shift on $j:$
$$ f_j=f(q_{j-1},q_j,q_{j+1}),\;
   g_j=g(q_{j-k},q_{j-k,x},\dots,q_{j+k},q_{j+k,x}).$$
Expanding this relation one obtains
\begin{eqnarray*}
&  {\partial f_1\over\partial q_{0}}D_t(F) +
   {\partial f_1\over\partial q_{1}}g_1 +
   {\partial f_1\over\partial q_{2}}g_2 =   &   \\
&= {\partial g_1\over\partial q_{1-k}}f_{1-k}        +
   {\partial g_1\over\partial q_{1-k,x}}D_x(f_{1-k}) +\dots +
   {\partial g_1\over\partial q_{1+k}}f_{1+k} +
   {\partial g_1\over\partial q_{1+k,x}}D_x(f_{1+k}) + R &
\end{eqnarray*}
where $R$ contains all terms with the second order partial derivatives of
$g_1.$  It is easy to see that the left hand side of the last equality does
not depend on the variables $q_{-k},q_{-k,x},$ therefore the term
$ {\partial g_1\over\partial q_{1-k,x}}
  {\partial f_{1-k}\over\partial q_{-k}}q_{-k,x} $
in the right hand side is equal to 0, under the condition (\ref{bc}).  But
for $k>1$ it is obvious that
${\partial f_{1-k}\over\partial q_{-k}}\vert_{(\ref{bc})}\neq 0,$
therefore ${\partial g_1\over\partial q_{1-k,x}}\vert_{(\ref{bc})}=0.$
The factor ${\partial g_1\over\partial q_{1-k}}$ is also equal to 0, for
otherwise the first term in the right hand side of equality would essentially
depend on variable $q_{-k}.$

Thus, under condition (\ref{bc}) the function $g_1$ does not depend on
$q_{1-k}, q_{1-k,x}.$  Assembling factors at independent variables
$q_{1-k},q_{1-k,x}\dots,q_{-2},q_{-2,x}$ and repeating the reasoning, one
gets that
${\partial g_1\over\partial q_{-j}}  \vert_{(\ref{bc})}\equiv 0,\;
 {\partial g_1\over\partial q_{-j,x}}\vert_{(\ref{bc})}\equiv 0,\;
 j=1,2,\dots,k-1,$
i.e. the right hand side of the equation
$q_{1,t}=g(q_{1-k},q_{1-k,x},\dots,q_{1+k},q_{1+k,x})$
under condition (\ref{bc}) actually depends only on dynamic variables
$q_1,q_{1,x},q_2,q_{2,x},\dots.$  The fact that other equations of the
lattice (\ref{sym}) for $n>0$ do not depend on variables $q_{-j},q_{-j,x},$
if $q_0=F$ is checked similarly \qed

\th{Consequence.}  The degenerate boundary condition reduces commuting
infinite lattices (\ref{chain}),(\ref{sym}) into commuting semi-infinite
lattices, given on the half-axis $j>0.$

Notice that a similar fact is true for the lattices
$q_{j,x}=f(q_{j+1},q_j,q_{j-1})$ of the Volterra type as well.

\th{Definition.} The boundary problem (\ref{chain}),(\ref{bc}) is called
compatible with the higher symmetry (\ref{sym}), if one of the following
conditions is fulfilled:

i) Boundary condition (\ref{bc}) is degenerate;

ii) Semi-infinite lattice reduced from (\ref{sym}) by virtue of (\ref{bc})
and its differential consequences obtained by differentiating it with
respect to $x$ by virtue of (\ref{chain}) (see above algorithm of
continuation) commutes with the semi-infinite lattice
(\ref{chain}),(\ref{bc}).

One can associate with the pair of lattices (\ref{chain}),(\ref{sym}) a
system of two partial differential equations.  For this purpose one passes
from the standard set of dynamical variables $q_0,q_{0,x},q_{\pm 1},$ $q_{\pm
1,x},\dots$ to the dynamical set consisting of the variables $q_0, q_1$ and
their derivatives with respect to $x.$  Thus $q_{-1}$ and $q_2$ are expressed
from the conditions $q_{0,xx}=f(q_1,q_0,q_{-1})$ and
$q_{1,xx}=f(q_2,q_1,q_0)$ and so on.  Rewriting the lattice (\ref{sym}) in
the new variables, one comes to a system of evolutionary equations (cf.
\cite{sy})
\bb \begin{array}{l}
 q_{0,t}=g_+(q_0,q_1,q_{0,x},q_{1,x},\dots,q_{0,x\dots x},q_{1,x\dots x}) \\
 q_{1,t}=g_-(q_0,q_1,q_{0,x},q_{1,x},\dots,q_{0,x\dots x},q_{1,x\dots x})
    \end{array}
\label{ev}\ee
This transformation maps the boundary condition (\ref{bc}) into the
differential constraint of the form
\bb q_0=\hat F(q_1,q_{0,x},q_{1,x},\dots,q_{0,x\dots x},q_{1,x\dots x})
\label{consq}\ee
The following criterion of the boundary condition compatibility with the
higher symmetry is a direct consequence of the way the system (\ref{ev}) was
constructed.

\th{Proposition 2.3} In order that the boundary problem
(\ref{chain}),(\ref{bc}) be compatible with the symmetry (\ref{sym}), it is
necessary and sufficient that the differential constraint (\ref{consq}) be
consistent with the dynamics of the equation (\ref{ev}).

\section{Differential constraints compatible with the ZS-AKNS hierarchy}

Let us rewrite the higher symmetries of the Toda lattice
\bb q_{j,t_2}=q^2_{j,x} +\exp(q_{j+1}-q_j)+\exp(q_j-q_{j-1}),  \label{t2}\ee
\bb q_{j,t_3}=q^3_{j,x} +(2q_{j,x}+q_{j+1,x})\exp(q_{j+1}-q_j)
                        +(2q_{j,x}+q_{j-1,x})\exp(q_j-q_{j-1}) \label{t3}\ee
as the systems of partial differential equations, introducing new variables
$u_j=\exp q_{j+1},\; v_j=\exp(-q_j)$ and expressing $q_k, q_{k,x}$ through
derivatives of these variables with respect to $x$ owing to the Toda lattice
(see \cite{sy}).  By virtue of the lattice (\ref{t2}) each pair of the
variables $u_j, v_j$ satisfies the ZS-AKNS system
\bb u_{t_2}=u_{xx}+2u^2v, \;\; -v_{t_2}=v_{xx}+2v^2u     \label{akns}\ee
and by virtue of the lattice (\ref{t3}) --- the system of coupled KdV
equations
\bb u_{t_3}=u_{xxx}+6uvu_x, \;\; v_{t_3}=v_{xxx}+6vuv_x. \label{ckdv}\ee
Other higher symmetries of the Toda lattice also correspond to the higher
symmetries from ZS-AKNS hierarchy.  According to the proposition 2.3, finding
of a boundary condition of the form (\ref{bc}) compatible with some higher
symmetry of the Toda lattice is reduced to finding of differential constraint
of the form (\ref{consq}) compatible with dynamics, determined by this
symmetry.  Rewriting this constraint in variables $u_0,v_0$ and omitting the
subscript 0 for brevity, we come to the following problem.

\th{Problem.} Find the differential constraints of the form
\bb F(u,v,u_x,v_x,\dots, u_{x\dots x},v_{x\dots x}) =0  \label{consuv}\ee
compatible with the $k$-th symmetry of the ZS-AKNS hierarchy, that is those
satisfying the identity
\bb D_{t_k}(F)\Bigr\vert_{F=0}\equiv 0.                 \label{defF}\ee

Let us call the order of the differential constraint (\ref{consuv}) the
maximum order of derivatives with respect to $x$ contained in it.  Finding of
constraint of the given order compatible with a given flow is reduced to
direct, but rather tedious calculations.  However, it is easy to prove that
because of skew symmetry of the main part, the system (\ref{akns}) and its
even order symmetries do not admit any differential constraint, except
trivial ones
\bb u=0 \;\; \mbox{or} \;\; v=0,              \label{c0}\ee
which correspond to the boundary conditions $\exp(q_1)=0$ or $\exp(-q_0)=0.$
Thus, the Problem can have nontrivial solution only for the flows,
appropriate to the times $t_{2k+1}.$  Apparently, if the constraint
(\ref{consuv}) is compatible with some odd order higher symmetry, then it is
compatible with the system (\ref{ckdv}) as well.  Basing on this hypothesis,
we shall use the system (\ref{ckdv}) as a test symmetry when constructing
boundary conditions for the Toda lattice.  The following statements are
proven by direct computations.

\th{Proposition 3.1} The constraint $u=P(v)$ is compatible with dynamics of
the equation (\ref{ckdv}) in only case if $P$ is a linear function:
\bb u = \alpha v +\beta. \label{c1}\ee

\th{Proposition 3.2} The differential constraint of the first order is
compatible with the system (\ref{ckdv}) if and only if it is of the form
$u_x=a(u,v)v_x,$ where the factor $a(u,v)$ satisfies the Hopf equation
$a_v+aa_u=0.$

\th{Remark.} In essence, this constraint is reduced to the linear constraint
from the previous proposition.  Really, the function $a$ is the first
integral of relation $u_x=av_x;$ assuming $a=\alpha$ and integrating, one
obtains (\ref{c1}).

\th{Proposition 3.3} The differential constraint of the form
$u=P(v,v_x,v_{xx})$ compatible with the system (\ref{ckdv}) is set by the
formula
\bb u(v^2-c_0)={v^2_xv+c_2(v^2+c_0)+c_1v\over v^2-c_0}-v_{xx} \label{c2}\ee
where $c_0,c_1,c_2$ are arbitrary constants.

Finding of more complex constraints directly along the definition
(\ref{defF}) becomes difficult.  For further progress we need a method, which
would allow us to reproduce new examples from already found ones.  For this
purpose it is enough to find operations acting on the set of integrable
constraints.  Two such operations are obvious.  Indeed, the system
(\ref{ckdv}) is invariant under the transformations
\bb S_c:\;\; \ti u=cu,\; \ti v=v/c,\; c=\mbox{const}    \label{dil}\ee
\bb R:\;\;   \ti u=v,\;  \ti v=u                        \label{refl}\ee
hence rewriting the constraint (\ref{consuv}) in the new variables yields
integrable constraint again.  For example, in the formula (\ref{c2}) it is
possible to swop places of $u$ and $v.$  The less obvious operation is
application of the B\"acklund (auto-)transformations.  As is well-known (see,
e.g \cite{sy}) the system (\ref{ckdv}) admits two essentially different
B\"acklund transformations.  One of them is given by the explicit formula
\bb T:\;\; v_1=1/u,\; u_1= u_{xx} - u^2_x/u +u^2v       \label{shft}\ee
and is equivalent to the shift $q_j\rightarrow q_{j+1}$ in the Toda lattice.
Substituting expression (\ref{shft}) into the constraint (\ref{consuv})
imposed on the variables $u_1, v_1$ one finds a new constraint on the
variables $u,v.$  In this case the orders of both constraints differ not more
than by 2.  For example, from the constraint $u_1=1$ one obtains (using in
addition the reflection $R$) the constraint (\ref{c2}) with constants
$c_0=c_1=0,c_2=1.$

The second B\"acklund transformation (more precisely, its "$x$-part") is of
the form
\bb  B_\mu :\;\;      u_x =\ti u +\mu u +u^2\ti v, \;\;
                 -\ti v_x = v +\mu\ti v +\ti v^2u.      \label{bt}\ee
Differentiation of these relations gives expressions for $v$ and derivatives
of $u$ and $v$ of any order through $u$ and derivatives of $\ti u,\ti v.$
Substituting these expressions into the constraint (\ref{consuv}), one
obtains some relation of the form $\hat F(u,\ti u,\ti v,\dots,\ti u_{x\dots
x},\ti v_{x\dots x})=0.$ Eliminating the variable $u$ from the equations
$\hat F=0$ and $D_x(\hat F)=0$ one obtains for the variables $\ti u,\ti v$ a
differential constraint
$$ \ti F(\ti u,\ti v,\dots,\ti u_{x\dots x},\ti v_{x\dots x}) =0. $$
Since, according to the definition of the B\"acklund transformation, the new
variables $\ti u,\ti v$ also satisfy the system (\ref{ckdv}), hence the found
constraint appears compatible with this system as well.  We call the
described procedure dressing of the constraint (\ref{consuv}).  It is easy to
check that dressing also changes the order of constraint not more than by 2.

The indicated transformations generate some group with relations
\bb\begin{array}{c}
 S_cS_d=S_{cd},\; RS_cR=S_{1\over c},\; S_cT=TS_c,\; S_cB_\mu=B_\mu S_c, \\
 R^2=1,\; RTR=T^{-1},\; RB_\mu R = S_{-1}B^{-1}_{-\mu},\;
   TB_\mu =B_\mu T,\; B_\mu B_\nu =B_\nu B_\mu.
\end{array}\label{code}\ee
Our hypothesis is that all constraints compatible with the odd flows of the
ZS-AKNS hierarchy form the orbit of two simplest constraints $u=1$ and $u=0$
under action of this group.  Notice that for all constraints obtained in such
a way the consistence with the odd flows really takes place.  It follows from
the facts that these flows are compatible with the constraint
$u=\mbox{const}$ and that the transformations (\ref{dil})-(\ref{bt}) act on
all hierarchy.  The constraints obtained as a result of application of
several transformations $T$ and $B_\mu$ to the nondegenerate constraint $u=1$
are listed in the table 1.
Let us comment the formulae indicated in it.  Denoting result of $m$-th
iteration of the B\"acklund transformation through $u(m),v(m)$ one obtains
from (\ref{bt}) a lattice
\bb  u_x(m)=u(m+1) +\mu(m)u(m) +u^2(m)v(m+1),           \label{ux}\ee
\bb -v_x(m+1)=v(m) +\mu(m)v(m+1) +v^2(m+1)u(m).         \label{vx}\ee
It is obvious that by virtue of the constraint $u(0)=1$ the equation
(\ref{ux}) at $m=0$ turns into the constraint $u(1)+v(1)+\mu(0)=0$ of the
form (\ref{c1}).  Notice that variables $v(0),v(1)$ satisfy the equations KdV
and mKdV
$$ v_t(0)=v_{xxx}(0)+6v(0)v_x(0), \;\;
   v_t(1)=v_{xxx}(1)-6(v(1)+\mu(0))v(1)v_x(1), $$
respectively and the equation (\ref{vx}) at $m=0$ turns into the Miura map
$$ -v(0)=v_x(1)+\mu(0)v(1)+v^2(1) $$
between them.  On the next step the equation (\ref{vx}) at $m=1$ yields
\bb u(1)=-{v_x(2)+\mu(1)v(2)-\mu(0)\over v^2(2)-1} \label{cons12}\ee
and, substituting in (\ref{ux}) at $m=1$ one obtains, after simple
calculations, the constraint (\ref{c2}) with constants
$$ c_0=1,\; c_1=-\mu^2(0)-\mu^2(1),\; c_2=\mu(0)\mu(1) $$
written for the variables $u(2),v(2).$  Notice that using the transformation
(\ref{dil}) one can obtain constraint (\ref{c2}) with arbitrary choice of
parameters.  Variable $v(2)$ satisfies the equation
\bb v_t=v_{xxx}-{6vv_x\over v^2-1}
   \Bigl(v_{xx}-{vv^2_x+c_2(v^2+1)+c_1v\over v^2-1}\Bigr). \label{eq2}\ee
Equation (\ref{cons12}) can be rewritten in the form of differential
substitution
$$ v(1) = {v_x(2)+\mu(1)v(2)-\mu(0)v^2(2)\over v^2(2)-1} $$
connecting mKdV and equation (\ref{eq2}).  Notice, that the point
transformation $v=\tanh y$ brings (\ref{eq2}) into the so called exponential
Calogero-Degasperis equation \cite{cd}
$$ y_t = y_{xxx} -2y^3_x -{3\over 8}\bigl((\mu(1)-\mu(0))^2e^{4y}
         +(\mu(1)+\mu(0))^2e^{-4y} -2\mu^2(1)-2\mu^2(0)\bigl)y_x. $$
Further application of the transformation (\ref{bt}) results in too
cumbersome constraint of 4-th order.  Appears, however, that the combination
$T^{-1}B_{\mu(2)}$ gives constraint of the 2-nd order again.  It can
conveniently be written in the variables
$$ 2r=u_{-1}(3)-v_{-1}(3),\; 2s=u_{-1}(3)+v_{-1}(3). $$
Omitting rather bulky calculations we put only the answer:
\bb 16r^2(s^2_x+P)=(2s_{xx}+\dot P)^2           \label{c3}\ee
where
$$ P=(s-\alpha)(s-\alpha_0)(s-\alpha_1)(s-\alpha_2),\;
   2\alpha=\mu(0)+\mu(1)+\mu(2),\; \alpha_j=\mu(j)-\alpha. $$
Rewriting the system (\ref{ckdv}) in the new variables and eliminating $r$ by
virtue of the constraint one finds that variable $s$ satisfies the elliptic
Calogero-Degasperis equation
$$s_t=s_{xxx}+6\Bigl(s^2-{(2s_{xx}+\dot P)^2\over 16(s^2_x+P)}\Bigr)s_x$$
introduced in \cite{cd} as well.  The differential substitution connecting
this equation with equation (\ref{eq2}) is given by the formula
$$ (c_2+2\mu(2)s-2s_x)v^2 +(c_1+\mu(2)^2+4s^2)v +c_2+2\mu(2)s+2s_x=0. $$
Differential substitutions for the scalar evolution equations are in
detail described in the literature, see e.g \cite{ssy}.

\section{Boundary conditions for the Toda lattice}

Let us write down the boundary conditions corresponding to the differential
constraints (\ref{c0})-(\ref{c2}) found above (certainly the constraint
(\ref{c3}) gives some boundary condition as well, but we do not need it).
Passing to the variables $q_j$ one obtains
\bb \exp(q_1)=0 \quad\mbox{or}\quad \exp(-q_0)=0        \label{b0}\ee
\bb \exp(q_1)=\alpha\exp(-q_0)+\beta                    \label{b1}\ee
\bb c_0\exp(q_1)=\exp(-q_{-1})
     +{c_0q^2_{0,x}+c_2(c_0\exp(q_0)+\exp(-q_0))+c_1
      \over c_0\exp(q_0)-\exp(-q_0)}                    \label{b2}\ee
where $\alpha,\beta,c_0,c_1,c_2$ are arbitrary parameters.  It should be
mentioned that the boundary conditions (\ref{b0}),(\ref{b1}) were found
earlier in \cite{mos} and \cite{skl} respectively.

It is well known (see, e.g. \cite{lez},\cite{bog},\cite{op}) that each simple
Lie algebra of the finite growth corresponds to some integrable generalized
Toda lattice.  The lattices associated with the Lie algebras $A_{n-1}, B_n,
C_n$ are obtained by imposing of the degenerate boundary condition
$\exp(-q_0)=0$ on the left end of the lattice (\ref{toda}) and boundary
conditions (\ref{b0}), (\ref{b1}) of particular form on its right end, namely
$\exp(q_{n+1})=0,\; q_{n+1}=0,\; q_{n+1}=-q_n$ respectively.  Now we shall
demonstrate that lattices of the $D$ type correspond to the boundary
condition of the form (\ref{b2}).  Lie algebra $D_n$ corresponds to the
lattice
\bb \begin{array}{rcl}
 q_{1,xx}  & = & e^{q_2-q_1},                                           \\
 q_{j,xx}  & = & e^{q_{j+1}-q_j} - e^{q_j-q_{j-1}}, \;\; j=2,\dots,n-2, \\
 q_{n-1,xx}& = & (e^{q_n}+e^{-q_n})e^{-q_{n-1}} - e^{q_{n-1}-q_{n-2}},  \\
 q_{n,xx}  & = & (e^{-q_n}-e^{q_n})e^{-q_{n-1}}.
\end{array} \label{Dn}\ee
It is clear that the boundary condition on the left end is of the form
$e^{-q_0}=0,$ i.e. is degenerate.  In order to bring the last but one
equation of the system (\ref{Dn}) into the standard form we make the change
of variables
\bb \hat q_j=q_j-\log 2,\; j=1,\dots,n-1,\;\; e^{\hat q_n}=\cosh q_n.
\label{hatq}\ee
Thus the last equation of the system accepts the form (we omit the hat over
variables $q_j$)
$$ q_{n,xx}=(e^{-q_n}-e^{q_n})e^{-q_{n-1}}+{q^2_{n,x}\over e^{2q_n}-1}. $$
Assuming $q_{n,xx}=e^{q_{n+1}-q_n}-e^{q_n-q_{n-1}}$ one obtains a boundary
condition for the right end of the lattice
\bb e^{q_{n+1}}=e^{-q_{n-1}}+{q^2_{n,x}\over 2\sinh q_n}. \label{a}\ee
Obviously, up to the shift $q_0\mapsto q_n,$ the formula (\ref{a}) is a
particular case of (\ref{b2}) for $c_0=1,\; c_1=c_2=0.$  So, the Toda lattice
of the type $D_n$ is reduced, by use of the change of variables (\ref{hatq}),
to the following finite-dimensional reduction of the Toda lattice
(\ref{toda}):
\bb\begin{array}{rcl}
     e^{-q_0} & = & 0,                                                  \\
     q_{j,xx} & = & e^{q_{j+1}-q_j} - e^{q_j-q_{j-1}}, \;\; j=1,\dots,n \\
  e^{q_{n+1}} & = & e^{-q_{n-1}}+{q^2_{n,x}\over 2\sinh q_n}.
\end{array}\label{hatDn}\ee
Now we can represent systems related to the Lie algebras $A_{n-1},B_n,C_n,
D_n,BD_n,BA_{2n-2},$ $\ti B_{n-1},\ti C_{n-1},CA_{2n-3},\ti D_{n-1}$ (in
other words, only the series $\ti A$ corresponding to the periodic Toda
lattice and exceptional Lie algebras drop out our consideration) in the form
of the problem (\ref{toda})-(\ref{bound}) with boundary conditions of the
type (\ref{b0})-(\ref{b2}).  Actually in this matter structure of the
beginning and the end of the Lie algebra Coxeter-Dynkin diagram is important
only.  The left(right) column of the table 2 contains the boundary conditions
on the left(right) end of the lattice and corresponding variants of the
beginning(end) of the diagram.  All possible combinations of the boundary
conditions on the left and right ends give us the Toda lattices appropriate
to all Lie algebras listed above.

We shall demonstrate now that the series $D$ lattices can be obtained from
the series $A$ lattices as a result of reflection type reduction combined
with the B\"acklund transformation (\ref{bt}).  For this purpose we need to
rewrite the transformations (\ref{dil})-(\ref{bt}) in the variables $q_j:$
$$ S_c:\; \ti q_j = q_j+\log c,  \quad
   R:\;   \ti q_j = -q_{1-j},    \quad
   T:\;   \ti q_j = q_{j+1}, $$
\bb B_\mu:\;\; \left\{
  \begin{array}{ccl}
    \ti q_{j,x} & = & \exp(\ti q_j-q_j) +\exp(q_{j+1}-\ti q_j)+\mu, \\
        q_{j,x} & = & \exp(\ti q_j-q_j) +\exp(q_j-\ti q_{j-1})+\mu.
  \end{array} \right.
\label{tbt}\ee
The generalized Toda lattice corresponding to the Lie algebra $A_{m-1}$ is
obtained from the infinite lattice by imposing two degenerate boundary
conditions:
\bb\begin{array}{ccl}
 q_0      & = & +\infty,                                            \\
 q_{j,xx} & = & e^{q_{j+1}-q_j} - e^{q_j-q_{j-1}}, \;\; j=1,\dots,m \\
 q_{m+1}  & = & -\infty.
\end{array}\label{Am}\ee
It is easy to check that the formula (\ref{tbt}) complemented by the
conditions
\bb \ti q_0=q_0=+\infty,\;\; \ti q_{m+1}=q_{m+1}=-\infty  \label{btb}\ee
sets the B\"acklund transformation for the system (\ref{Am}).  Thus the
mapping $B_\mu: Q \mapsto \ti Q,$ where $Q=(q_1,q_{1,x},\dots,q_m,q_{m,x})$
is given by explicit formulae and is convertible.  Indeed, the relations
(\ref{tbt}),(\ref{btb}) represent a closed system of algebraic equations
relatively $\ti q_{j,x},\exp(\ti q_j),$ which can easily be solved
recurrently, starting from the equation $q_{1,x}=\exp(\ti q_1-q_1)+\mu.$
Notice, that the given transformation is an example of integrable discrete
mapping.  It can be interpeted as finite dimensional reduction of the twice
discrete lattice induced by transformation (\ref{tbt}).  About integrable
mappings see e.g \cite{brst},\cite{ves},\cite{pnc}.

The following statement establishes connection between lattices of the series
$A$ and $D.$

\th{Proposition 4.1} Let $m=2n$ and the solution $Q$ of the system (\ref{Am})
satisfies the reduction condition $Q=S_{-1}RT^mQ,$ that is
$q_j=i\pi-q_{m+1-j}.$  Then the new solution $\ti Q=B_0Q$ of the system
(\ref{Am}) possesses the following properties:

1) $\ti Q = B^2_0 RT^m\ti Q,$

2) First half of the vector $\ti Q,$ that is the vector
   $(\ti q_1,\ti q_{1,x},\dots,\ti q_n,\ti q_{n,x})$
   solves the lattice (\ref{hatDn}).

\proof The property 1) can easily be proven with the help of relations
(\ref{code}).  Further, from reduction conditions on $q$ it follows that
$q_{n+1}=i\pi-q_n,$ that is $u_n=-v_n.$  It follows from the table 1 that
dressing of this constraint by transformation $B_0$ yields the constraint of
the form (\ref{c2}) with constants $c_0=1,\;c_1=c_2=0$ connecting the
variables $\ti u_n,\ti v_n.$  As we have already seen this constraint is
equivalent to the boundary condition (\ref{a}). \qed

Concluding this section we note that the similar simple connection can be
established between the type $A$ lattices and the lattices with the boundary
condition (\ref{b2}) of general form on the right end.  The proof of the
following proposal does not cause difficulties.

\th{Proposition 4.2} Let $m=2n+1$ and the solution $Q$ of the system
(\ref{Am}) satisfies the reduction condition $q_j=-q_{m+1-j}.$  Then the new
solutions of this system $\ti Q= S_cB_\mu Q,\; \bar Q=B_\nu\ti Q$ where
$c,\mu,\nu$ are arbitrary parameters solve the Toda lattices with the
boundary condition $\exp(-q_0)=0$ on the left end and boundary conditions
(\ref{b1}),(\ref{b2}) of the general form on the right end respectively.

It follows from the well known fact on the general solution of the type $A_m$
lattice (see, e.g. \cite{lez}) and propositions 4.1 and 4.2 that the general
solutions of such lattices can be written explicitly as rational functions on
the exponents $e^{\lambda x}.$

\section{Boundary conditions with explicit dependence on $x$.}

The boundary conditions found above admit some generalization.  Notice that
the Toda lattice (\ref{toda}) is invariant under transformation
\bb X:\;\; \ti q_j = q_j+\ve x,                         \label{xq}\ee
or, in the variables $u,v$
\bb X:\;\; \ti u = e^{\ve x}u,\;\; \ti v= e^{-\ve x}v.  \label{xuv}\ee
The difference between this transformation and transformations
$S_c,R,T,B_\mu$ considered earlier is that it does not preserve the higher
symmetries of the Toda lattice. For example, the system (\ref{ckdv}) at this
change of variables is rewritten as follows
\begin{eqnarray*}
 & u_{t_3}=u_{xxx}+6uvu_x -3\ve(u_{xx}+2u^2v) +3\ve^2u_x -\ve^3u, & \\
 & v_{t_3}=v_{xxx}+6vuv_x +3\ve(v_{xx}+2v^2u) +3\ve^2v_x +\ve^3v. &
\end{eqnarray*}
In general, it is easy to check that change (\ref{xuv}) determines some
linear transformation in the space of symmetries of the ZS-AKNS hierarchy.
Boundary conditions found in the previous section are consistent with the odd
order symmetries subspace.  If this subspace were invariant under the
transformation $X$ then rewriting of these boundary conditions in variables
$\ti q_j$ would give nothing new.  The presence of the even order terms in
the transformed symmetries results in that the transformed boundary
conditions contain explicit dependence on $x$.  For example, the formulae
(\ref{b1}), (\ref{b2}) are converted into the formulae
\bb \exp(q_1)=\alpha\exp(2\ve x-q_0)+\beta\exp(\ve x),  \label{xb1}\ee
\bb c_0\exp(q_1)=\exp(2\ve x-q_{-1})
     +{c_0(q_{0,x}-\ve)^2+c_2(c_0\exp(q_0-\ve x)+\exp(\ve x-q_0))+c_1
      \over c_0\exp(q_0-2\ve x)-\exp(-q_0)}             \label{xb2}\ee
correspondingly.  It should be emphasized, that new boundary conditions are
not worse than old ones, but they are compatible with other symmetry set.  At
imposing on both ends of the Toda lattice of boundary conditions relevant to
the same value $\ve$ this symmetry set passes into the symmetries of obtained
finite-dimensional system and provides its integrability.  The situation,
however, essentially changes if the boundary conditions imposed on the
different ends of the lattice correspond to the different values of parameter
$\ve.$  It turns out that in small dimensions the equations of the Painlev\'e
type arise (compare with \cite{ad}, where the Painlev\'e equations arise at
quasiperiodic closing of integrable lattices).  In the examples below the
length 1 reductions of the Toda lattice are considered.  On the left end we,
without loss of generality, impose boundary conditions of the type (\ref{b1})
or (\ref{b2}) and on the right one deformed conditions (\ref{xb1}) or
(\ref{xb2}) at some fixed value of $\ve.$  (It is clear that the boundary
condition (\ref{b0}) will not give anything new, since the transformation
(\ref{xq}) does not change it.)

1) Let us try boundary conditions of the form (\ref{b1}) and (\ref{xb1}) at
$\ve=2$:
$$  e^{-q_{-1}}=\alpha e^{q_0}+\beta,\;\;
    e^{q_1} =\gamma e^{4x-q_0}+\delta e^{2x}. $$
On the variable $q=q_0$ one obtains the differential equation
$$ q_{xx}=\gamma e^{4x-2q} +\delta e^{2x-q} -\alpha e^{2q} -\beta e^q.$$
It is easy to check that the point transformation
$e^{q(x)}=zy(z),\;\;e^x=z$ brings it into the 3-rd Painlev\'e equation
$$y_{zz}={y^2_z\over y}-{y_z\over z}+{1\over z}(Ay^2+B)+Cy^3+{D\over y}$$
with the values of parameters
$A=-\beta,\;B=\delta,\;C=-\alpha,\;D=\gamma.$

2) Impose the boundary condition of the form (\ref{b2}) with $c_0=1$ without
loosing of generality on the left end and a boundary condition (\ref{xb1})
with $\ve=1$ on the right one:
$$ e^{-q_{-1}} = e^{q_1} -
    {q^2_{0,x}+c_2(e^{q_0}+e^{-q_0})+c_1 \over e^{q_0}-e^{-q_0}}, \;\;
 e^{q_1}=\alpha e^{2x-q_0}+\beta e^x. $$
On the variable $q=q_0$ one obtains the differential equation
$$ q_{xx} = {q^2_x\over 1-e^{-2q}} +\alpha e^{2x}(e^{-2q}-1)
    +\beta e^x(e^{-q}-e^q) +{c_2(e^q+e^{-q})+c_1\over 1-e^{-2q}}. $$
It is easy to check that the point transformation
$e^{q(x)}={y(z)-1\over y(z)+1},\;\;e^x=z$
brings it into the 5-th Painlev\'e equation
$$ y_{zz} = ({1\over 2y}+{1\over y-1})y^2_z -{y_z\over z}
  +{(y-1)^2\over z^2}(Ay+{B\over y}) +C{y\over z} +D{y(y+1)\over y-1} $$
with the values of parameters
$8A=-c_1-2c_2,\;8B=c_1-2c_2,\;C=2\beta,\;D=2\alpha.$

3) At last, consider a combination of boundary conditions of the form
(\ref{b2}) and (\ref{xb2}) (thus, because of small dimension of the system,
the left and right ends actually coincide).  It is easy to prove that from 7
parameters, contained in these two formulas, 3 can be replaced by any nonzero
constants by use of dilating $x,$ shift $x$ and shift $q.$  Therefore we may
consider boundary conditions
$$ e^{q_1} = e^{-q_{-1}} +{q^2_{0,x} +a(e^{q_0}+e^{-q_0})+b\over
                                        e^{q_0}-e^{-q_0}},\;\;
   e^{q_1} = e^{2x-q_{-1}} +{(q_{0,x}-1)^2 +c(e^{q_0-x}+e^{x-q_0})+d\over
                                        e^{q_0-2x}-e^{-q_0}}. $$
Solving these equations relatively $e^{q_1},e^{-q_{-1}} $ and substituting
found values into the formula $q_{0,xx}=e^{q_1-q_0}-e^{q_0-q_{-1}},$ one
obtains some differential equation of the 2-nd order on the variable $q=q_0.$
One can check up that change of variables
$ e^{q(x)}={y(z)+\sqrt{z}\over y(z)-\sqrt{z}},\;\;
       e^x={1+\sqrt{z}\over 1-\sqrt{z}}$
converts it into the 6-th Painlev\'e equation
\begin{eqnarray*}
 y_{zz} &=& {1\over 2}({1\over y}+{1\over y-1}+{1\over y-z})y^2_z
                     -({1\over z}+{1\over z-1}+{1\over y-z})y_z + \\
        & & +{y(y-1)(y-z)\over z^2(z-1)^2}
  (A +B{z\over y^2} +C{z-1\over (y-1)^2} +D{z(z-1)\over (y-z)^2})
\end{eqnarray*}
with parameters
$ 8A=-b-2a,\; 8B=b-2a,\; 8C=-d-2c,\; 8D=d-2c+4.$

\section*{Acknowledgements}

The authors thank A.B. Shabat, S.I. Svinolupov, V.V. Sokolov and R.I. Yamilov
for interest to work and useful discussions.  This work was partially
supported by Russian Foundation of Fundamental Researches (grants
93-011-16088 and 93-011-165) and International Scientific Foundation (grants
MLY000 and RK-2000).


\begin{figure}[p]
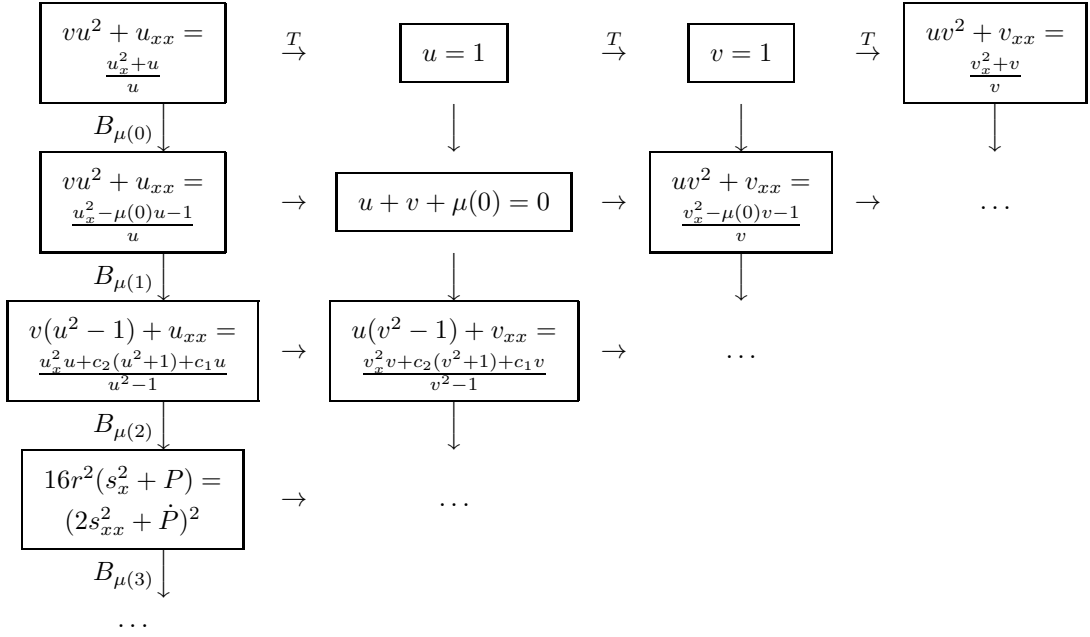
 
$$ \begin{array}{ccccccc}
 \dbox{vu^2+u_{xx}=}{{u^2_x+u\over u}} & \stackrel{T}{\rarrow} &
 \dbox{u=1}{}                          & \stackrel{T}{\rarrow} &
 \dbox{v=1}{}                          & \stackrel{T}{\rarrow} &
 \dbox{uv^2+v_{xx}=}{{v^2_x+v\over v}}                                    \\
          B_{\mu(0)} \darrow & & \darrow  & & \darrow  & & \darrow        \\
 \dbox{vu^2+u_{xx}=}{{u^2_x-\mu(0) u-1\over u}} & \rarrow &
 \dbox{u+v+\mu(0)=0}{}                          & \rarrow &
 \dbox{uv^2+v_{xx}=}{{v^2_x-\mu(0) v-1\over v}} & \rarrow & \dots         \\
          B_{\mu(1)} \darrow & & \darrow  & & \darrow  & &                \\
 \dbox{v(u^2-1)+u_{xx}=}{{u^2_xu+c_2(u^2+1)+c_1u\over u^2-1}} & \rarrow &
 \dbox{u(v^2-1)+v_{xx}=}{{v^2_xv+c_2(v^2+1)+c_1v\over v^2-1}} & \rarrow &
 \dots & &                                                                \\
          B_{\mu(2)} \darrow & & \darrow  & &          & &                \\
 \dbox{16r^2(s^2_x+P)=}{(2s^2_{xx}+\dot P)^2} & \rarrow & \dots & & & &   \\
          B_{\mu(3)} \darrow & &          & &          & &                \\
 \dots & & & & & &
\end{array}
$$
\caption{Dressing of the differential constraints}
\end{figure}

\pagebreak

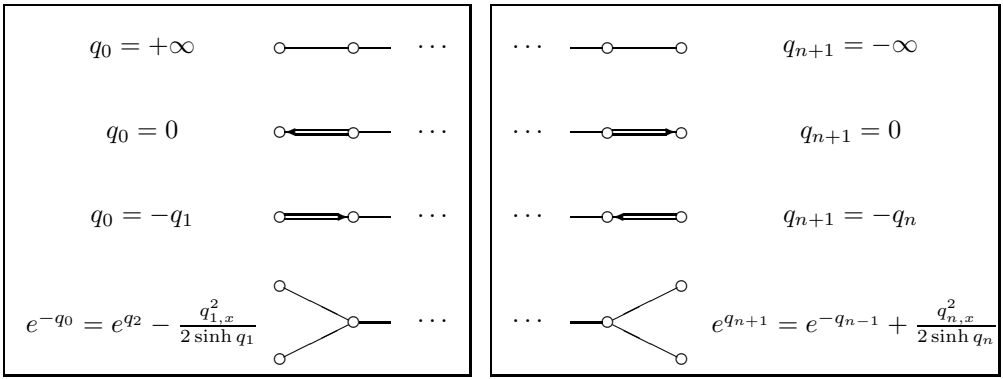
\begin{figure}[p] 
 \setlength{\unitlength}{0.04em}
 \begin{picture}(900,360)(-30,0)
   \put(120,310){\makebox(0,0){$q_0=+\infty$}}
   \put(250,310){\circle{10}}    \put(255,310){\line(1,0){60}}
   \put(320,310){\circle{10}}    \put(325,310){\line(1,0){30}}
   \put(380,310){$\dots$}

   \put(120,230){\makebox(0,0){$q_0=0$}}
   \put(250,230){\circle{10}}    \put(260,230){\vector(-1,0){5}}
   \put(265,232){\line(1,0){50}} \put(265,228){\line(1,0){50}}
   \put(320,230){\circle{10}}    \put(325,230){\line(1,0){30}}
   \put(380,230){$\dots$}

   \put(120,150){\makebox(0,0){$q_0=-q_1$}}
   \put(250,150){\circle{10}}    \put(310,150){\vector(1,0){5}}
   \put(255,152){\line(1,0){50}} \put(255,148){\line(1,0){50}}
   \put(320,150){\circle{10}}    \put(325,150){\line(1,0){30}}
   \put(380,150){$\dots$}

   \put(120,50){\makebox(0,0)
        {$e^{-q_0}=e^{q_2}-{q^2_{1,x}\over 2\sinh q_1}$}}
   \put(250,85){\circle{10}}    \put(250,15){\circle{10}}
   \put(254,83){\line(2,-1){61}}\put(254,17){\line(2,1){61}}
   \put(320,50){\circle{10}}    \put(325,50){\line(1,0){30}}
   \put(380,50){$\dots$}

   \put(-10,0){\framebox(440,350){}}
   \put(450,0){\framebox(480,350){}}

   \put(790,310){\makebox(0,0){$q_{n+1}=-\infty$}}
   \put(630,310){\circle{10}}    \put(625,310){\line(-1,0){60}}
   \put(560,310){\circle{10}}    \put(555,310){\line(-1,0){30}}
   \put(470,310){$\dots$}

   \put(790,230){\makebox(0,0){$q_{n+1}=0$}}
   \put(630,230){\circle{10}}    \put(620,230){\vector(1,0){5}}
   \put(615,232){\line(-1,0){50}}\put(615,228){\line(-1,0){50}}
   \put(560,230){\circle{10}}    \put(555,230){\line(-1,0){30}}
   \put(470,230){$\dots$}

   \put(790,150){\makebox(0,0){$q_{n+1}=-q_n$}}
   \put(630,150){\circle{10}}    \put(570,150){\vector(-1,0){5}}
   \put(625,152){\line(-1,0){50}}\put(625,148){\line(-1,0){50}}
   \put(560,150){\circle{10}}    \put(555,150){\line(-1,0){30}}
   \put(470,150){$\dots$}

   \put(790,50){\makebox(0,0){
        $e^{q_{n+1}}=e^{-q_{n-1}}+{q^2_{n,x}\over 2\sinh q_n}$}}
   \put(630,85){\circle{10}}     \put(630,15){\circle{10}}
   \put(626,83){\line(-2,-1){61}}\put(626,17){\line(-2,1){61}}
   \put(560,50){\circle{10}}     \put(555,50){\line(-1,0){30}}
   \put(470,50){$\dots$}

\end{picture}
\caption{Boundary conditions corresponding to the ends of Coxeter-Dynkin
         diagrams}
\end{figure}

\end{document}